\documentclass[twocolumn,printnumbers,amsmath,amssymb,prl]{revtex4}
\usepackage{graphicx}
\usepackage{color}

\begin{document}

\title{Rheological similarities between dense self-propelled and sheared particulate systems}

\author{Ruoyang Mo}
\author{Qinyi Liao}
\author{Ning Xu$^*$}

\affiliation{Hefei National Laboratory for Physical Sciences at the Microscale, CAS Key Laboratory of Microscale Magnetic Resonance and Department of Physics, University of Science and Technology of China, Hefei 230026, P. R. China\\$^*$ningxu@ustc.edu.cn}

\begin{abstract}
Different from previous modelings of self-propelled particles, we develop a method to propel the particles with a constant average velocity instead of a constant force. This constant propulsion velocity (CPV) approach is validated by its agreement with the conventional constant propulsion force (CPF) approach in the flowing regime. However, the CPV approach shows its advantage of accessing quasistatic flows of yield stress fluids with a vanishing propulsion velocity, while the CPF approach is usually unable to because of finite system size. Taking this advantage, we realize the cyclic self-propulsion and study the evolution of the propulsion force with propelled particle displacement, both in the quasistatic flow regime. By mapping shear stress and shear rate to propulsion force and propulsion velocity, we find similar rheological behaviors of self-propelled systems to sheared systems, including the yield force gap between the CPF and CPV approaches, propulsion force overshoot, reversible-irreversible transition under cyclic propulsion, and propulsion bands in plastic flows. These similarities suggest the underlying connections between self-propulsion and shear, although they act on systems in different ways.
\end{abstract}

\maketitle

\section{\label{sec:level1}Introduction}

When moving together, living objects or the so-called active matter, e.g., birds, fish, and bacteria, usually exhibit complex collective behaviors. One of the special features of active matter is self-propulsion, i.e., it consumes energy and propels itself to move.  Because of self-propulsion, the collective motion of active matter is inherently non-equilibrium, leading to rich and unusual dynamical and phase behaviors \cite{ramaswamy2017arcmp,marchetti2013rmp,bechinger2016rmp,reichhardt2017arcmp,ramaswamy2017jstat}.

In previous modelings, active matter is simplified to self-propelled particles (SPPs) performing the run-and-tumble motion \cite{marchetti2016minimal,fily2012athermal}. At any instant, each SPP is propelled by a force in a randomly chosen direction. The particle tumbles when the direction of the force changes. Therefore, a system consisting of a large number of SPPs is driven to flow by forces randomly distributed in space. The fascinating aspect is that the SPPs can self organize and form clustering or flocking patterns \cite{marchetti2016minimal,henkes2011active,bialke2015active,fily2012athermal,berthier2014nonequilibrium}.

Viewed from another perspective, self-propulsion is just one of the ways to activate the system. In this aspect, it would be natural to consider comparing it with shear, another typical and widely employed driving mechanism. Taking planar shear flows as example, when a shear force ${\bf F}_{\rm s}=F_{\rm s}\hat{\bf x}$ is applied in parallel to the top surface of a system, it results in a shear stress $\Sigma=F_{\rm s}/S$ and a shear rate $\dot\gamma={v_{\rm s}/L}$ in steady state, where $S$ is the area of the surface, $L$ is the length of the system in the direction perpendicular to the surface, $v_{\rm s}$ is the velocity of the surface in the direction of the unit vector $\hat{\bf x}$. Similarly, in a steady flow of a self-propelled system, the propulsion force ${\bf f}_i=f_{i}\hat{\bf n}_i$ acting on SPP $i$ results in a propulsion velocity $v_i$ in the direction of the unit vector $\hat{\bf n}_i$. Averaged over all SPPs, there are an average propulsion force and an average propulsion velocity in the directions of $\{\hat{\bf n}_i\}$ ($i=1,2,...,N$) with $N$ being the total number of SPPs. Qualitatively, if propulsion force plays a similar role as shear stress (or shear force), propulsion velocity would be plausibly the counterpart of shear rate (or velocity of the shearing surface). Of course, the differences between self-propulsion and shear are obvious: Self-propulsion is local (microscopic) and random in space, while shear is global (macroscopic) and directional. It is thus not so straightforward to make a direct connection between them. However, our recent study has shown a sign of such a connection \cite{liao2018criticality}. By mapping shear stress and shear rate to propulsion force and propulsion velocity, we found similar critical scalings in the vicinity of the jamming transition \cite{liu1998jamming,o2003jamming,liu2010arcmp,hecke2010jp} between self-propelled and sheared systems \cite{liao2018criticality,liu2014finite,olsson2007critical}. Whilst most of the studies of active matter attempt to find new phenomena, the thinking of whether different non-equilibrium systems share similarities is apparently lacking \cite{agoritsas}. This is however of fundamental significance in developing the theory of non-equilibrium systems.

In this work, we concentrate on the high-density regime where particles form amorphous solids in the absence of self-propulsion or shear, aiming at finding out whether self-propulsion can cause similar responses of solids to shear. The mechanical response or rheology of amorphous solids subject to shear has been widely studied and some special features have been observed \cite{jaiswal2016mechanical,tanguy2006plastic,rottler2003shear,ozawa2018random,jiang2015origin,lin2015criticality,langer2001microstructural,leishangthem2017yielding,regev2013onset,parmar2019strain,priezjev2017collective,fiocco2013oscillatory}. Here we list some typical phenomena which will be studied and compared for self-propelled systems. First, under quasistatic shear, an amorphous solid undergoes elastic deformation with the shear stress increasing with the shear strain, until yielding and reaching the steady state of plastic flows, in which the shear stress fluctuates around the yield stress. It has been shown that for well annealed amorphous solids there is a stress overshoot prior to yielding \cite{ozawa2018random,jiang2015origin,lin2015criticality,langer2001microstructural}. Second, under cyclic shear, an amorphous solid undergoes the reversible-irreversible transition \cite{leishangthem2017yielding,regev2013onset,parmar2019strain}. When the amplitude of the cyclic shear, $\gamma_{\rm a}$, is smaller than a critical value $\gamma_{\rm a,m}$, the system can eventually reach a reversible state whose potential energy decreases with the increase of $\gamma_{\rm a}$. When $\gamma_{\rm a}>\gamma_{\rm a,m}$, the system cannot be reversible any more. Third, it has been found that shear bands form in plastic flows \cite{priezjev2017collective,ozawa2018random,fiocco2013oscillatory}.

Note that shear strain is the control parameter in all the studies listed above. For systems consisting of SPPs, it is equivalent to demanding the control of propelled particle displacement. To simply the problem, in this study we only consider the minimal model of SPPs without tumbling, i.e., the direction of self-propulsion on each particle remains fixed. In previous modelings, SPPs are thus driven by a constant force, apparently not good for the control of particle displacement. It is then required to develop a new model with the propulsion velocity being under control. Here we construct such a constant propulsion velocity (CPV) model. The model is validated by the agreement of its flow curve, i.e., propulsion force versus propulsion velocity, with that of the conventional constant propulsion force (CPF) approach. With the CPV model, we are able to access quasistatic flows of amorphous solids driven by self-propulsion. We do observe similar rheological features to sheared systems as mentioned above, when shear stress and shear rate (strain) are replaced with propulsion force and propulsion velocity (particle displacement). Therefore, although acting on particles in rather different ways, self-propulsion and shear result in similar rheological behaviors of amorphous solids, suggesting some underlying connections between them.

\section{\label{sec:level1}Methods}

Our systems are boxes with a side length $L$ and periodic boundary conditions in all directions. To avoid crystallization, we employ binary mixtures of $N/2$ large and $N/2$ small SPPs. The diameter ratio of large to small SPPs is $1.4$. SPPs $i$ and $j$ interact via the purely repulsive harmonic potential
\begin{equation}
U(r_{ij}) = \frac{\epsilon}{2}\left(1-\frac{r_{ij}}{\sigma_{ij}}\right)^2\Theta\left(1-\frac{r_{ij}}{\sigma_{ij}}\right), \label{pot}
\end{equation}
where $\epsilon$ is the characteristic energy scale of the interaction, $r_{ij}$ and $\sigma_{ij}$ are the separation between SPPs $i$ and $j$ and the sum of their radii, and $\Theta(x)$ is the Heaviside step function.

The movement of SPP $i$ is governed by the equation of motion
\begin{equation}
\chi\frac{{\rm d} \textbf{r}_i}{{\rm d} t}=\textbf{F}_i + f\hat{\textbf{n}}_i=- \sum_{j}\frac{{\rm d} U(r_{ij})}{{\rm d} \textbf{r}_i}+f \hat{\textbf{n}}_i, \label{eom}
\end{equation}
where $\textbf{r}_i$ is the location of particle $i$, $\chi$ is the damping coefficient, $f$ and $\hat{\textbf{n}}_i$ are the magnitude and direction of the self-propulsion force, and the sum is over all SPPs interacting with SPP $i$, resulting in a force $\textbf{F}_i$. As usually done in simulations of SPPs, we employ overdamped particle dynamics, so that the inertia of SPPs is ignored. Here we consider the minimal model with $\hat{\textbf{n}}_i$ being fixed in time or switching between $\hat{\textbf{n}}_i$ and $-\hat{\textbf{n}}_i$ when studying cyclic self-propulsion, so SPPs do not tumble randomly. In the beginning of the simulation, we randomly assign each SPP a self-propulsion direction and adjust them to ensure $\sum_i{\hat{\textbf{n}}}_i=0$, so that there is no net force on the system.

In the conventional CPF approach, $f$ in Eq.~(\ref{eom}) remains constant. When particle interactions are absent, SPP $i$ will move in the direction of $\hat{\textbf{n}}_i$ with a constant velocity $f/\chi$. However, when particles interact, the velocities of SPPs will change in time. As discussed above, the CPF approach is analogous to the constant shear stress approach of sheared systems. We thus need a CPV approach analogous to the constant shear rate approach, with the average self-propulsion velocity or the average particle displacement in the direction of self-propulsion being the control parameter.

The CPV approach can be straightforwardly achieved by the dot product of $\hat{\textbf{n}}_i$ and Eq.~(\ref{eom}):
\begin{equation}
\chi \textbf{v}_i\cdot \hat{\textbf{n}}_i=\textbf{F}_i\cdot \hat{\textbf{n}}_i + f,
\end{equation}
where $\textbf{v}_i=\frac{{\rm d}\textbf{r}_i}{{\rm d}t}$ is the velocity of SPP $i$. By defining $\frac{1}{N}\sum_{i=1}^N\textbf{v}_i\cdot\hat{\textbf{n}}_i=v$, we have
\begin{equation}
f = \chi v - \frac{1}{N}\sum_{i=1}^N \textbf{F}_i\cdot\hat{\textbf{n}}_i. \label{cv}
\end{equation}
Then, for a given average self-propulsion velocity $v$, the self-propulsion force $f$ is no longer constant. It fluctuates in time and is determined by Eq.~(\ref{cv}). Combining Eqs.~(\ref{eom}) and (\ref{cv}), we realize the CPV approach with the propulsion velocity $v$ being under control.

Here we only show results for two-dimensional systems. The major findings reported in the following sections should be valid for three-dimensional systems as well. We set the units of energy and length to be $\epsilon$ and small particle diameter $\sigma$. The time is in units of $\chi \sigma^2/ \epsilon$. The temperature is in units of $\epsilon/k_{\rm B}$ with $k_{\rm B}$ being the Boltzmann constant. For sheared systems discussed and studied in this work, we mainly concern about simple shear.

\section{Comparing CPV and CPF approaches}

A system can be driven to flow when a constant shear stress (CSS) $\Sigma$ or a constant shear rate (CSR) $\dot\gamma$ is applied. In steady state, the shear stress fluctuates around an average value when the shear rate is fixed, and vice versa. In the flowing regime, it has been shown that the flow curves $\Sigma(\dot\gamma)$ of the CSS and CSR approaches agree \cite{xu2006measurements}. However, for yield stress fluids which have a nonzero shear stress $\Sigma_{\rm yv}$ in the $\dot\gamma=0$ limit, it has also been shown that the system under CSS can cease flowing even when the applied shear stress is still larger than $\Sigma_{\rm yv}$ \cite{xu2006measurements}. If we define another yield stress $\Sigma_{\rm yf}$ as the smallest shear stress for the system to flow forever from the CSS perspective, $\Sigma_{\rm yf}>\Sigma_{\rm yv}$ for finite size systems \cite{varnik2003shear,varnik2004study,xu2006measurements}. With the increase of system size, $\Sigma_{\rm yf}$ approaches $\Sigma_{\rm yv}$. It is expected that $\Sigma_{\rm yf}=\Sigma_{\rm yv}$ in the thermodynamic limit \cite{xu2006measurements}.

For self-propelled systems, the CPV and CPF approaches should be analogous to the CSR and CSS approaches for sheared systems, respectively. It is thus necessary to check first whether the flow curves $f(v)$ of the CPV and CPF approaches can match. For yield stress fluids, there should also exist a yield force $f_{\rm yv}$ in the $v=0$ limit. It is interesting to know whether the yield force $f_{\rm yf}$ defined from the CPF perspective \cite{liao2018criticality} is greater than $f_{\rm yv}$ and approaches $f_{\rm yv}$ when system size increases.

\begin{figure}
	\includegraphics[width=0.45\textwidth]{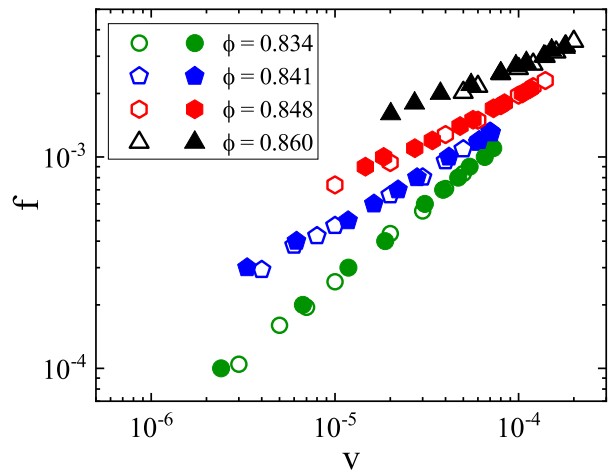}
	\caption{Comparison of the flow curves $f(v)$ of the $N=1024$ systems for CPV and CPF approaches at various packing fractions $\phi$. The solid and empty symbols are for CPF and CPV approaches, respectively. }
	\label{fig:fig1}
\end{figure}

In both CPV and CPF approaches, we start the simulation from random states and collect data after the systems reach steady state. Figure~\ref{fig:fig1} compares the flow curves $f(v)$ of the two approaches at different packing fractions $\phi$. They agree very well, confirming that the CPV approach works.

\begin{figure}[ht!]
\vspace{-0.05in}
	\includegraphics[width=0.5\textwidth]{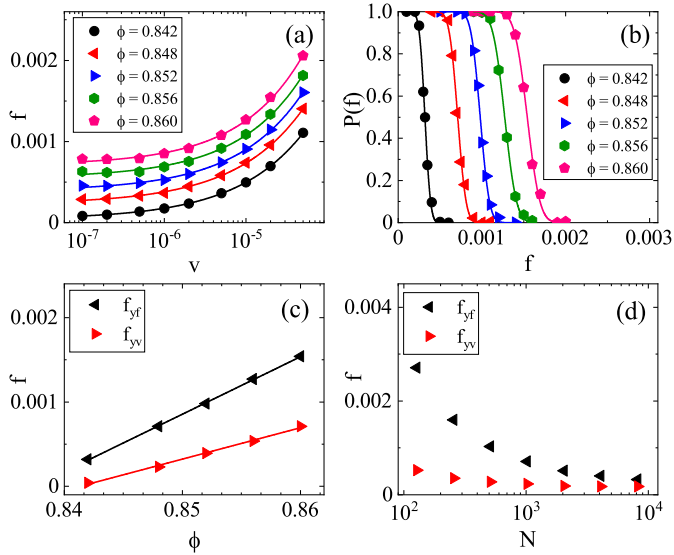}
	\caption{Comparison of the yield forces $f_{\rm yv}$ and $f_{\rm yf}$ from the CPV and CPF approaches. (a) Flow curves $f(v)$ of the $N=1024$ systems obtained from the CPV approach. The lines are fits to the Herschel-Bulkley law. The fitting parameters of Eq.~(\ref{eq:HB}) are $f_{\rm yv}=(4.18\times 10^{-5}, 2.30\times 10^{-4}, 3.92\times 10^{-4}, 5.54\times 10^{-4}, 7.11\times 10^{-4})$, $A_{\rm v}=(0.195, 0.216, 0.258, 0.296, 0.361)$, and $\alpha=(0.526, 0.526, 0.541, 0.550, 0.564)$ for $\phi=(0.842, 0.848, 0.852, 0.856, 0.860)$, respectively. (b) Probability of finding jammed states $P(f)$ for the $N=1024$ systems obtained from the CPF approach with the stopping criterion of the search being $\Delta r(t) \geq L$, as discussed in the text. The lines are fits to the error function. (c) Packing fraction $\phi$ dependence of $f_{\rm yv}$ and $f_{\rm yf}$ for the $N=1024$ systems. The lines are linear fits. (d) System size $N$ dependence of $f_{\rm yv}$ and $f_{\rm yf}$.}
	\label{fig:fig2}
\end{figure}

In Fig.~\ref{fig:fig2}(a), we focus on the $v\rightarrow 0$ part of the $f(v)$ curves obtained from the CPV approach. As will be shown later, this regime is inaccessible by the CPF approach, indicating the advantage of the CPV approach. The $f(v)$ curves tend to approach a constant in the $v=0$ limit and can be fitted well with the Herschel-Bulkley law \cite{HB}:
\begin{equation}
f = f_{\rm yv} + A_{\rm v} v^\alpha,\label{eq:HB}
\end{equation}
where $f_{\rm yv}$ is the yield force from the CPV perspective, and $A_{\rm v}$ and $\alpha$ are fitting parameters. It is well-known that $\Sigma(\dot\gamma)$ of shear flows can also be fitted well with the Herschel-Bulkley law \cite{mueller,besseling,engmann,otsuki,dinkgreve,paredes}. As shown in the caption of Fig.~{\ref{fig:fig2}}, $A_{\rm v}$ and $\alpha$ vary with packing fraction. For two-dimensional shear flows of underdamped passive particles interacting with harmonic potential, it has been shown that $\alpha\approx 0.65$ at $\phi=0.85$ \cite{xu2006measurements}, slightly different from the values shown here. Therefore, although the flow curves can all be fitted well with the Herschel-Bulkley law, the values of fitting parameters still depend on properties of systems, e.g., materials, density, and dynamics \cite{otsuki,dinkgreve,paredes}.

In the CPF approach, we determine the yield force $f_{\rm yf}$ from the probability of finding jammed states (solids), $P(f)$ \cite{liu2014finite,liao2018criticality,zheng2016shear}. Because the system is athermal, when $f$ is small, it can quickly converge to a jammed state and cease flowing. When $f$ is large, the system always flows within the accessible time window. For finite size systems, $P(f)$ decays from $1$ to $0$ with the increase of $f$. Figure~\ref{fig:fig2}(b) shows examples of $P(f)$ at different packing fractions. Here we stop the search of jammed states when the average particle displacement exceeds the cutoff $L$: $\Delta r(t)=N^{-1} \sum_{i}|\textbf{r}_i(t)-\textbf{r}_i(0)| \geq L$, where the sum is over all SPPs \cite{liao2018criticality}. At a given packing fraction $\phi$, $P(f)$ can be fitted well with the error function: $P(f)=\frac{1}{2}{\rm erfc}\left( \frac{f- f_{\rm yf}}{\sqrt{2}w} \right)$, where $w$ measures the variance or the uncertainty of the yield force $f_{\rm yf}$. As shown in our previous studies \cite{liu2014finite,liao2018criticality}, $w$ decreases with the increase of system size and tends to vanish in the thermodynamic limit. For finite size systems, the yield force $f_{\rm yf}$ defined here is the force at which the probability of finding jammed states is $50\%$.

As discussed in our previous study \cite{liao2018criticality}, $P(f)$ increases if $\Delta r(t)$ is allowed to reach a larger cutoff. This is simply because that the system flowing at the cutoff $L$ may finally find a jammed state to resist the force $f$ and get stuck there when $\Delta r(t)$ is allowed to reach $nL$ ($n=2,3,...$). Therefore, $f_{\rm yf}$ extracted from Fig.~\ref{fig:fig2}(b) is smaller than the ``ideal" yield force in the infinite waiting time limit. However, it is impractical to wait forever. It has been shown that the growth of $P(f)$ with the increase of $n$ follows a recurrence relation \cite{liao2018criticality}, so it is reasonable to use the same stopping criterion to estimate $f_{\rm yf}$, as done for Fig.~\ref{fig:fig2}(b).

Figure~\ref{fig:fig2}(c) compares $f_{\rm yv}(\phi)$ with $f_{\rm yf}(\phi)$. Even though $f_{\rm yf}$ is smaller than the ``ideal" value in the infinite waiting time limit, it is greater than $f_{\rm yv}$, similar to the fact that $\Sigma_{\rm yf}>\Sigma_{\rm yv}$ for sheared systems \cite{xu2006measurements}. Moreover, Fig.~\ref{fig:fig2}(c) shows that both $f_{\rm yv}(\phi)$ and $f_{\rm yf}(\phi)$ are linear and tend to vanish at $\phi_{\rm Jv}$ and $\phi_{\rm Jf}$, respectively, in agreement with the linear packing fraction dependence of the yield stresses for systems with harmonic repulsion \cite{liu2014finite,liao2018criticality}. Here $\phi_{\rm Jv}$ and $\phi_{\rm Jf}$ are expected jamming transition threshold in the $f\rightarrow 0$ limit viewed from CPV and CPF approaches, respectively.

Figure~\ref{fig:fig2}(d) shows the system size dependence of $f_{\rm yv}$ and $f_{\rm yf}$ at a given packing fraction. When system size increases, both yield forces decrease, leading to the decrease of the gap between them. In the large system size limit, $f_{\rm yf}$ and $f_{\rm yv}$ show the trend to converge to the same value. This again reminisces the relationship between $\Sigma_{\rm yf}$ and $\Sigma_{\rm yv}$ for sheared systems \cite{xu2006measurements}. Similar to the fact that the system size dependence of $\Sigma_{\rm yf}$ is stronger than that of $\Sigma_{yv}$ for sheared systems \cite{xu2006measurements,liu2014finite}, Fig.~\ref{fig:fig2}(d) shows that the variation of $f_{\rm yf}$ with system size is also stronger than that of $f_{\rm yv}$, which is responsible to the discrepancy between $\phi_{\rm Jv}$ and $\phi_{\rm Jf}$ extracted from Fig.~\ref{fig:fig2}(c). It is expected that $\phi_{\rm Jv}$ and $\phi_{\rm Jf}$ converge to the same value in the thermodynamic limit.

In this section, we validate the CPV approach and show the first evidence of the connections between self-propelled and sheared systems. The CPV approach is crucial to access quasistatic flows of amorphous solids under self-propulsion. Because $f_{\rm yf}>f_{\rm yv}$ for finite size systems and $f(v)$ curves of CPV and CPF approaches match, the self-propulsion velocity $v_{\rm yf}$ at $f=f_{\rm yf}$ is greater than $0$. Then, applying any $f\in (f_{\rm yv},f_{\rm yf})$ will eventually lead to a jammed state, so it is impossible for the CPF approach to drive the system to flow forever with an average self-propulsion velocity $v<v_{\rm yf}$.

Employing the CPV approach, we are able to attack the issues to be discussed next, which are all investigated in the quasistatic-flow regime. In the following sections, we always apply the CPV approach with $v=10^{-6}$. As shown in Fig.~\ref{fig:fig2}(a), this velocity is roughly in the quasistatic regime where the propulsion force already develops a plateau.

\begin{figure}
	\includegraphics[width=0.45\textwidth]{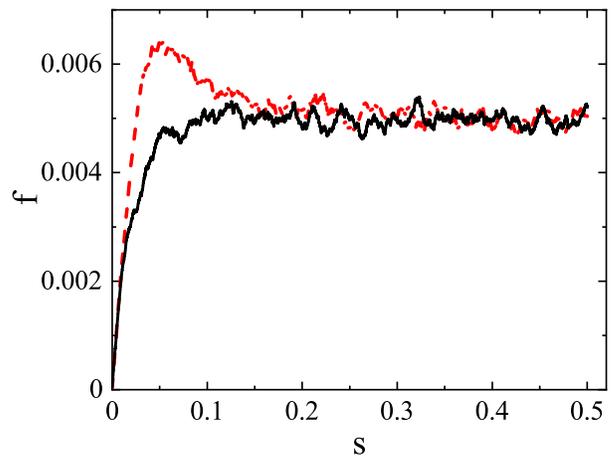}
	\caption{Comparison of the propulsion force $f$ versus average particle displacement $s$ along the direction of self-propulsion for quickly (solid) and slowly (dashed) quenched amorphous solids with $N=2048$ and $\phi=0.95$.}
	\label{fig:fig3}
\end{figure}

\section{Propulsion force overshoot}

When subject to quasistatic shear, an amorphous solid first undergoes the elastic deformation with the shear stress increasing with the increase of shear strain, and then experiences plastic flows with the shear stress fluctuating around the yield stress $\Sigma_{\rm yv}$. In certain circumstances, the shear stress exhibits an overshoot before reaching the plastic-flow regime \cite{jiang2015origin,ozawa2018random}. It has been shown that the emergence of the stress overshoot depends on the history of preparing the solid. If the solid is quickly quenched from a high-temperature state, there is usually no overshoot. In contrast, a slowly-quenched amorphous solid will show an overshoot \cite{ozawa2018random}. With the decrease of quench rate, the resultant amorphous solid will be more and more stable and some microscopic order may be established. However, such structural changes are not dramatic. Although the stress overshoot seems to be associated with the stability or microscopic ordering of the solid, a widely accepted understanding has not been well-established yet \cite{ozawa2018random,jiang2015origin,lin2015criticality,langer2001microstructural}.

As suggested by the previous work \cite{liao2018criticality} and the results shown in the previous section, propulsion force and propulsion velocity for self-propelled systems play qualitatively similar roles of shear stress and shear rate, respectively. It is then straightforward to ask whether the propulsion force can exhibit an overshoot under quasistatic self-propulsion, whose emergence also relies on quench rate.

Figure~\ref{fig:fig3} shows examples of $f(s)$, where $s=vt$ is the average particle displacement in the direction of self-propulsion. Here we compare results of quickly and slowly quenched solids. The quickly quenched solids are obtained by minimizing the potential energy of high-temperature states via the fast inertial relaxation engine (FIRE) method \cite{bitzek2006structural}. To obtain slowly quenched solids, we anneal the system from the liquid state with a small quench rate ${\rm d}T/{\rm d}t=10^{-6}$, where $T$ is the temperature. For quickly quenched solids, there is no overshoot in $f(s)$. In sharp contrast, $f(s)$ of slowly quenched solids shows a pronounced overshoot. The emergence of propulsion force overshoot for well-annealed amorphous solids is the second evidence of the similarities between self-propelled and sheared systems.

\section{Cyclic self-propulsion}

\begin{figure}
	\includegraphics[width=0.45\textwidth]{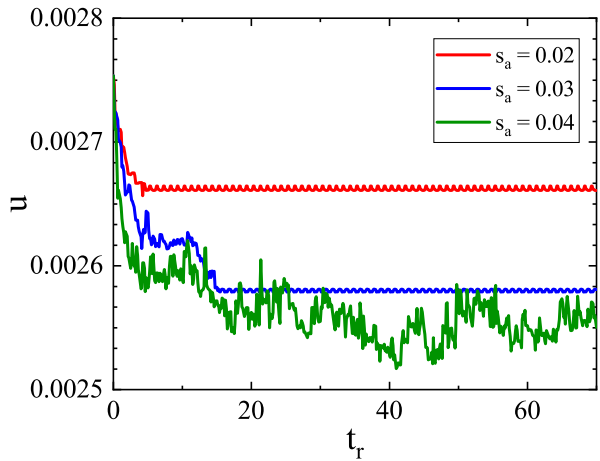}
	\caption{Potential energy per particle $u$ versus reduced time $t_{\rm r}=vt/(4s_{\rm a})$ for systems with $N=2048$ and $\phi=0.95$ under cyclic self-propulsion with different amplitude $s_{\rm a}$. }
	\label{fig:fig4}
\end{figure}

Recent studies have shown that under cyclic shear particulate systems such as colloidal suspensions and granular materials undergo the reversible-irreversible transition, upon the increase of the amplitude of the oscillatory shear \cite{fiocco2013oscillatory,pine2005chaos,regev2013onset}. Under a tiny shear strain amplitude, an amorphous solid deforms elastically. It is thus expected that the solid will return to its initial state after a cycle. When the amplitude increases, the solid may already undergo irreversible plastic rearrangements during the first cycle and cannot return to its initial state. However, if the amplitude is not large enough, a more stable state which can resist a large shear deformation without experiencing irreversible particle rearrangements may be found after some cycles. Then the system finally becomes reversible again. With the increase of the amplitude, the number of cycles required to find reversible states increases. There exists a critical amplitude above which the system cannot find reversible states any more within the accessible time window (or number of cycles).

Here we show that the reversible-irreversible transition exactly occurs for amorphous solids under cyclic self-propulsion, being the third evidence of the similarities between self-propelled and sheared systems. Starting with an amorphous solid, we let all SPPs move with an average self-propulsion velocity $v$ until the average particle displacement $s=vt$ in the direction of self-propulsion reaches the amplitude $s_{\rm a}$, and then reverse the direction of the self-propulsion until $s=-s_{\rm a}$. Afterwards, the direction of the self-propulsion is reversed again and a cycle is finished when $s$ returns to $0$.

\begin{figure}
	\includegraphics[width=0.45\textwidth]{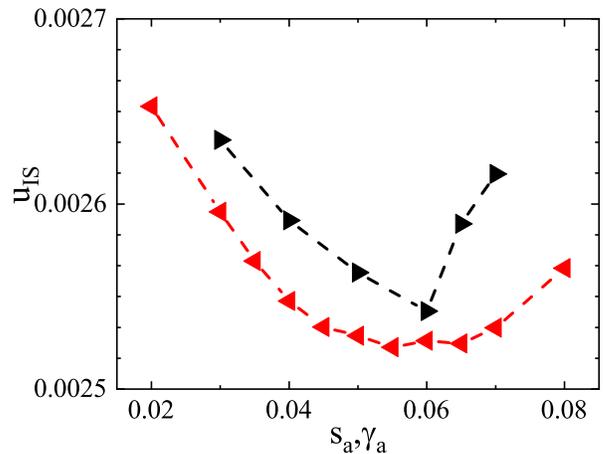}
	\caption{Potential energy per particle of the inherent structures minimized from the $s=0$ states of the last cycle, $u_{\rm IS}$, versus the amplitude of the cyclic self propulsion, $s_{\rm a}$ (red left triangles) for the $N=2048$ systems at $\phi=0.95$. For comparison, $u_{\rm IS}$ versus the amplitude of the cyclic shear, $\gamma_{\rm a}$ (black right triangles) is also shown. The lines are guides for the eye.}
	\label{fig:fig5}
\end{figure}

Figure~\ref{fig:fig4} shows the potential energy per particle, $u=\frac{1}{N}\sum_{ij}U(r_{ij})$, as a function of reduced time $t_{\rm r}=vt/(4s_{\rm a})$ for three values of $s_{\rm a}$, where the sum is over all interacting pairs of SPPs. For the smallest $s_{\rm a}$ shown here, the potential energy becomes repetitive after only several cycles. For the intermediate $s_{\rm a}$, it takes the system much more cycles to find the reversible state. For the largest $s_{\rm a}$, the system becomes irreversible within the time window of our simulation. There exists a critical $s_{\rm a}$ denoted as $s_{\rm a,m}$ at which the reversible-irreversible transition occurs.

Figure~\ref{fig:fig4} also shows that with the increase of $s_{\rm a}$ the potential energy of the reversible states decreases. This is expected because the reversible state found with a larger $s_{\rm a}$ can endure a larger deformation and is thus more stable with a lower potential energy. Therefore, the reversible state found with $s_{\rm a,m}$ should have the lowest potential energy. To verify that, we minimize the potential energy of the $s=0$ states in the $2000$th cycle via the FIRE method to get inherent structures, i.e., amorphous solids at zero temperature. Figure~\ref{fig:fig5} shows the potential energy $u_{\rm IS}$ averaged over tens of minimized inherent structures as a function of $s_{\rm a}$. The potential energy indeed reaches the minimum at a critical $s_{\rm a}$, which locates $s_{\rm a,m}$.

In Fig.~\ref{fig:fig5}, we also present results of cyclic quasistatic shear for comparison. To mimic quasistatic shear, we successively apply a small step strain of $10^{-4}$ to the system with Lees-Edwards boundary conditions \cite{allen_book}, followed by the potential energy minimization via the FIRE method. Cyclic quasistatic shear is realized by reversing the direction of shear when the strain amplitude $\gamma_{\rm a}$ is reached. The potential energy of the inherent structures shows a minimum as well. Therefore, both cyclic self-propulsion and shear are effective ways to search for more stable amorphous solids, which may be even helpful to the search of ultra-stable glasses \cite{ultra2007science,ninarello2017models}.

The comparison in Fig.~\ref{fig:fig5} may also suggest an interesting issue to be attacked. In contrast to $u_{\rm IS}(\gamma_{\rm a})$ which grows rapidly when $\gamma_{\rm a}>\gamma_{\rm a, m}$, the change of $u_{\rm IS}(s_{\rm a})$ is more gentle. It is interesting to know whether this reflects some intrinsic difference between self-propulsion and shear or is caused by some simulation parameters such as system size and propulsion velocity. When propulsion velocity $v$ decreases, the kinetic energy (temperature) of the system decreases. Analogous to the dependence of $u_{\rm IS}$ on the parent temperature of supercooled liquids \cite{ozawa2018random}, it would be expected that the decrease of propulsion velocity $v$ leads to the decrease of $u_{\rm IS}$. Here we use $v=10^{-6}$, which is small and roughly in the quasistatic regime, so a further decrease of $v$ may not significantly change the shape of $u_{\rm IS}(s_{\rm a})$. This requires more work to verify. In comparison, it may be more interesting to know how $u_{\rm IS}(s_{\rm a})$ and $u_{\rm IS}(\gamma_{\rm a})$ evolve with system size, in order to reveal the nature of the reversible-irreversible transition and identify whether self-propulsion and shear are really different in this aspect.

\section{Propulsion band in flocking pattern}

In plastic shear flows, it has been shown that shear bands can be formed by fast-moving particles \cite{priezjev2017collective,parmar2019strain,fiocco2013oscillatory}. The similarities between self-propelled and sheared systems already shown above stimulate us to check whether propulsion bands or clusters can also be formed in plastic self-propelled flows, which will then serve as the fourth evidence of the underlying connections between self-propelled and sheared systems.

\begin{figure}
	\includegraphics[width=0.49\textwidth]{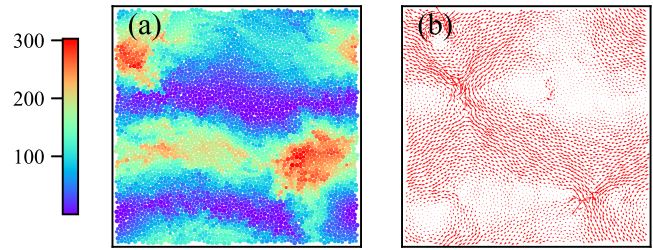}
	\caption{(a) Snapshot of a quasistatically self-propelled $N=4096$ system at $\phi=0.95$ in the plastic flow regime. The color on each SPP shows its squared particle displacement over a time interval of $4\times10^6$, as quantified by the color bar. (b) Velocity field of the same configuration in (a). The direction and length of the arrows show the direction and speed of the velocities of the SPPs.}
	\label{fig:fig6}
\end{figure}

In steady state of quasistatic flows of SPPs, we calculate the squared particle displacement $\Delta r_i^2(t) = |\textbf{r}_i(t)-\textbf{r}_{i}(0)|^2$ for SPP $i$ between two snapshots separated by a time interval of $t$. Figure~\ref{fig:fig6}(a) shows the spatial distribution of the squared particle displacement with a time interval of $4\times10^6$. Particles with large displacements aggregate and form a pronounced band or cluster, reminiscing the shear bands in quasistatic shear flows.

Figure~\ref{fig:fig6}(b) shows the corresponding velocity field in the configuration at the reference time $t=0$ when the particle positions start to be recorded. The direction and length of the arrows show the direction and speed of the velocities of the SPPs. The velocity field is nicely correlated with the spatial distribution of the particle displacement, with fast-moving SPPs forming a band overlapping well with the propulsion band in Fig.~\ref{fig:fig6}(a). Interestingly, the SPPs in the band tend to move in directions similar to their neighbors, organizing into a flocking pattern, although initially the SPPs are propelled in a totally random way.

\section{Discussion and Conclusions}

In summary, we develop a CPV approach to simulate self-propelled systems and apply it to study the rheology of amorphous solids. The CPV approach is validated by the agreement of its flow curves with those of the conventional CPF approach. For finite size systems, the CPV approach shows its advantage over the CPF approach of being able to explore quasistatic flows. Employing the CPV approach, we find four pieces of evidence reminiscing the rheological behaviors of amorphous solids under shear, suggesting the underlying connections between self-propelled and sheared systems. First, the yield force defined from the CPF approach is larger than that from the CPV approach. The gap between the two yield forces decreases with the increase of system size. Second, for well-annealed amorphous solids, there is an overshoot in the plot of the propulsion force against the particle displacement along the direction of self-propulsion before entering the plastic flow regime. Third, there is a reversible-irreversible transition under cyclic self-propulsion. Fourth, fast-moving SPPs form propulsion band and move in the flocking pattern.

Sheared and self-propelled systems are both typical non-equilibrium systems. The rheological similarities reported in this work suggest that there may be a unified framework to understand the two seemingly different systems with rather distinct loading mechanisms. In this work, we reveal their connections in the perspective of the rheology of amorphous solids in the quasistatic-flow regime. In this regime, both shear and self-propulsion act as the driving forces to deform the solids, so the responses observed here may just reflect the properties of the solids. In this aspect, it may be too early to conclude that sheared and self-propelled systems are intrinsically similar. More studies are definitely required to have a comprehensive comparison between the two systems, especially in the low-density and high-velocity regimes not considered in this work. It is also interesting to investigate in follow-up studies how the rheology will be affected when self-propulsion and shear work together.

\section{Acknowledgements}

This work is supported by National Natural Science Foundation of China Grant No. 11734014. We also thank the Supercomputing Center of University of Science and Technology of China for the computer time.

\bibliography{reference}

\end{document}